\documentclass[10pt,conference]{IEEEtran}

\usepackage{graphicx}

\begin{document}

\title{Analytical Performance Model for Poisson Wireless Networks with Pathloss and Shadowing Propagation}

\makeatletter
\let\thanks\@IEEESAVECMDthanks%
\makeatother

\author{
\authorblockN{
Jean-Marc Kelif$^1$,\thanks{$^1$Jean-Marc Kelif is with Orange Labs, France} 
\thanks{Email: jeanmarc.kelif@orange.com}
Stephane Senecal$^2$,\thanks{$^2$Stephane Senecal is with Orange Labs, France} 
\thanks{Email: stephane.senecal@orange.com} 
Marceau Coupechoux$^3$, \thanks{$^3$Marceau Coupechoux is with Telecom ParisTech, France} \thanks{Email: marceau.coupechoux@telecom-paristech.fr}
Constant Bridon$^4$ \thanks{$^4$Constant Bridon is with ENS Cachan and Orange Labs, France} 
\thanks{Email: constant.bridon@ens-cachan.fr}
}}


\maketitle

\begin{abstract}

The SINR (signal to interference plus noise ratio) is a key factor for wireless networks analysis. Indeed, the SINR distribution allows the derivation of performance and quality of service (QoS) evaluation. Moreover, it also enables the analysis of radio resources allocation and scheduling policies, since they depend on the SINR reached by a UE (User Equipment). Therefore, it is particularly interesting to develop an analytical method which allows to evaluate the SINR, in a simple and quick way, for a realistic environment.  
Considering a stochastic Poisson network model, we establish the CDF (cumulative distributed function) of the SINR. We show that the shadowing can be neglected, in many cases, as long as mobiles are connected to their best serving base station (BS), \textit{i.e.} the BS which offers them the most powerful useful signal. 
As a consequence, the analysis of performance and quality of service, directly derived from the CDF of SINR, can be established by using a propagation model which takes into account only the pathloss. Moreover, we establish that the Fluid network model we have proposed can be used to analyze stochastic Poisson distributed network. Therefore, the analysis of stochastic Poisson network can be done in an easy and quick way, by using the analytical expression of the SINR established thanks to the Fluid network model. 

\end{abstract}

%

\maketitle

\section{Introduction}

Mobile services demand is becoming more and more important. Therefore, the
estimation of performance and quality of service (QoS) has to be more and more
precise in the way to answer with accuracy to the demand. 
Performance and quality of service evaluations of
wireless networks can be analyzed by using simulations or
analytical models. 
Several QoS parameters (like throughput,
outage probability) can be derived from the SINR distribution \cite{Lagr05} \cite{Gil91} \cite{Ela05} \cite{Vit94}.

Moreover the knowledge of the signal to interference plus noise ratio (SINR) reached by a UE (User Equipment) allows to better perform the radio resource allocation and the scheduling policy. 
Analytical models thus try to derive simple SINR formula
in order to quickly evaluate the performance of a cellular
network. Therefore, their analysis need tractable and accurate
models of networks.
Two factors play an important role in the evaluation of the SINR :  the localization of BS and the propagation phenomena. 
Among the network model usually considered, the Hexagonal one is the most used. However this model is based on a regular deployment of BS on an area.  
The Fluid model, which is another model of network, considers the interfering base stations as a continuum of
infinitesimal interferers distributed in space. 
The main interest of this model consists in its tractability, in the possibility
to establish closed form formula of the SINR, whatever the
location of a UE, and to establish the SINR distribution \cite{KelCoPhycom11} \cite{KeSene12}, too. 

In this paper we consider a Poisson model network: the base stations are randomly distributed according to a spatial Poisson process \cite{Bac03} \cite{AgKeCoGo08}. This model allows to take into account more realistic environments than a hexagonal model of network. Indeed,  the distances between base stations are not constant. 
The propagation is generally modeled by a term which depends on the distance from the transmitter, the pathloss. Since in a real network, the power received at any point
of the system also depends on the local area, another term which characterizes local area impact can be added. The last term, the shadowing, is generally expressed as a lognormal distributed function \cite{KeCoGo08}.
Papers generally analyze the shadowing impact by considering users connected to their nearest BS \cite{KeCoGo08} \cite{KeCo09}.
Recent papers \cite{MatCoKe12} \cite{BarKar11} analyze the impact of shadowing by considering users connected to the BS which offers the highest useful signal.

\underline{Our contribution}: 
This paper focuses on the impact of the pathloss and the shadowing on the CDF of the SINR, considering that users are connected to their best serving station : UEs are connected to the BS which offers the highest useful signal.
We establish that in a Poisson network the CDF of the SINR calculated by considering a propagation model which takes into account the pathloss and the shadowing is very close to the CDF obtained without considering the shadowing (only the pathloss is considered). This result is very interesting. Indeed, it allows to use the simple expression of the SINR given by the Fluid network model. Therefore, the determination of the quality of service and performance of a wireless network can be done quickly and in a simple way.

The organization of the paper is as follows. In Section~\ref{systemmodel}, we present the system model. We recall, in Section \ref{FluidNetwork}, that performance and QoS of Poisson model can be established by using a Fluid model when the shadowing is not taken into account.  Section \ref{SINRCalculation} expresses the best server SINR. In Section \ref{shadowing}, the impact of shadowing is analyzed, considering UEs connected to the BS which offers the highest useful signal. In Section \ref{shadowingimpactsimulation}, some precisions are given about the impact of the analysis we developed on the study of realistic wireless systems. A conclusion is given in Section \ref{Conclusion}.

\section{System Model} \label{systemmodel}

Let us consider a wireless network. We focus on the downlink transmission part. Our aim is to evaluate the performance and the quality of service of a single user. We consider an access technology in which the radio resources of a base station (BS) are divided in a number of parallel, orthogonal, non-interfering channels (subcarriers), i.e. OFDMA. Therefore, only inter-cell interference is considered, no intra-cell interference.

\subsection{Network Topology : Poisson Network}
\begin{figure}[!h]
\centering
\includegraphics[height=5cm]{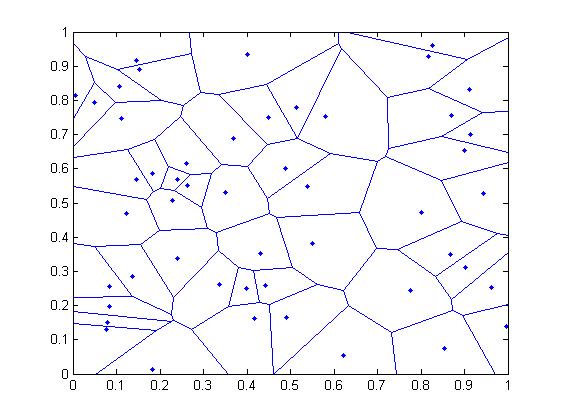}
\caption{Poisson Network} 
\label{Poissonnetwork}
\end{figure}

In a real network, the inter site distance is variable. The Poisson model network, characterized by the density of BS, allows to take it into account by considering a Poisson distribution of base stations in a given area (Fig. \ref{Poissonnetwork}). In this configuration, the cells of the network form a Vorono\"i diagram. Therefore, it becomes necessary to analyze a wide zone, with a great number of base stations, to determine the statistical characteristics of the network in terms of performance and quality of service.

\subsection{SINR of a user} 
We consider a single frequency network composed of $N$ base stations, transmitting at power $P$ on each subcarrier. We define $g_i(u)$ the path gain between BS $i$ and user $u$ on a given subcarrier. The SINR $\gamma_u$ of user $u$ served by BS $i$ on the considered subcarrier is given by:
\begin{equation} \label{SINR}
\gamma_u=\frac{P g_i(u) }{
\sum\limits_{j\neq i} P g_{j}(u)  + N_{th}}.
\end{equation}
with $ N_{th}$ the thermal noise on a subcarrier.

\subsection{Performance and quality of service} 
The knowledge of the SINR allows to calculate the throughput that may be reached by a user. Indeed, considering any subcarrier as an AWGN (Additive White Gaussian Noise) channel, the SINR received by a mobile enables the determination of the spectral efficiency $D_u$ (in bits/s/Hz) by using the Shannon formula: 

\begin{equation} \label{shannon}
D_u= \log_2(1+ \gamma_u). 
\end{equation}

Let us notice that there are alternative approaches like using a modified upper bounded Shannon formula or throughput-SINR tables coming from physical layer simulations. 
Moreover, expressions (\ref{SINR}) and (\ref{shannon}), calculated at any location of the network, allow an evaluation of the CDF (Cumulative Distribution Function) of the SINR (or the throughput). The SINR CDF also provides the outage probability, i.e. the probability that a user cannot be accepted in the network since he cannot have a sufficient throughput.
It is therefore important to develop a method which allows to determine these characteristics with a high accuracy, for a user at any distance $r$ from his serving BS.

Moreover, since the throughput allows to know the quality of service that can be offered to a user, these methods make it possible to determine this characteristic with a high accuracy in a simple way. In particular, the minimum throughput, obtained at cell edge, can be derived. By doing an integration all over the cell range, the average throughput of the cell can be calculated, too.
Dynamical analysis also need the knowledge of the SINR \cite{Rong11} as input.

\section{Reminder about the Fluid Network}\label{FluidNetwork}
\subsection{Fluid Network}
\begin{figure}[!h]
\centering
\includegraphics[height=5cm]{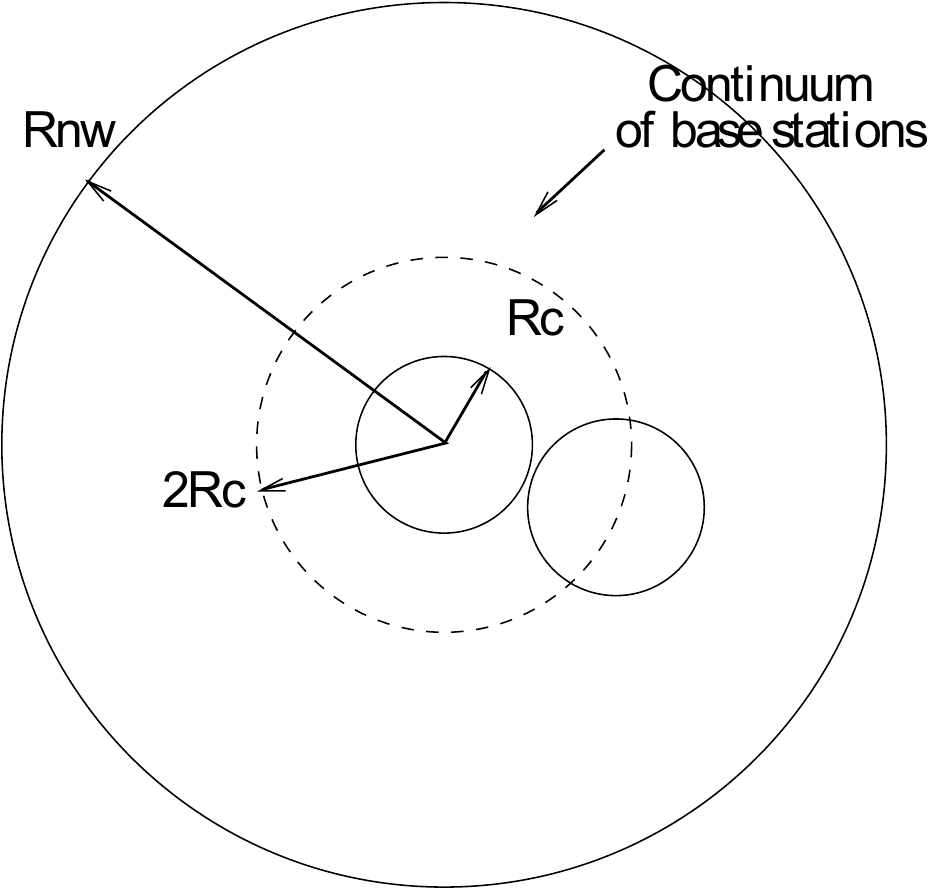}
\caption{Fluid model: Network and cell of interest} 
\label{Fluidnetwork}
\end{figure}

The fluid network model consists in replacing a given fixed finite number of transmitters by an equivalent continuous density of transmitters \cite{KeA05} \cite{KelCoEurasip10}. Given an inter site distance $2R_c$, 
interferers are characterized by a  density $\rho_{BS}$
of BS starting at a distance of $2R_c$ from a BS (covering a zone of radius $R_c$), as illustrated on Fig. \ref{Fluidnetwork} ($R_{nw}$
is the size of the network).
The interest of this model is to establish a simple analytical expression of the SINR.

\subsection{Calculation of the SINR}

We consider a path gain $g_{j}(u) = K r_j^{-\eta}(u)$, where  $K$ is a constant, $r_j(u)$ is the distance between user $u$ and BS $j$ and $\eta$ the path loss exponent. Let us consider that a BS (eNode-B) transmits at power $P$ on each subcarrier. Denoting r = $r_i$, we can express (\ref{SINR}) as (dropping u):

\begin{equation} \label{SIRdist}
\gamma=\frac{r^{-\eta} }{
\sum\limits_{j\neq i} r_j^{-\eta}},
\end{equation}
We consider a urban environment, where the thermal noise can be neglected.

By considering expression  (\ref{SIRdist}), we can see that the calculation of the analytical expression of the SINR of a user depends on its location, on the location of its serving BS, and on the location of each interfering BS. 

Considering the fluid model, the SINR only depends on the distance $r$ of the user to its serving base station \cite{Mas11}: 

\begin{equation}
\gamma(r)= \frac{\eta -2}{2\pi \rho_{BS}} \frac{r^{-\eta}}{(2R_c-r)^{2-\eta}}  \label{SIRfluid}
\end{equation}

If $\rho_{BS}$ is known, this analytical model allows to significantly simplify the calculation of the SINR: the only required variable is the distance of the mobile to its serving BS. This model has been proven to be reliable and close to the reality for homogeneous hexagonal networks \cite{KeA05} \cite{KelCoEurasip10}, as well as for heterogeneous networks \cite{KeSene12}.

The cell edge throughput can particularly be calculated by setting $r$ = $R_c$ in (\ref{shannon}) and   (\ref{SIRfluid}). Therefore, the minimum performance and quality of service offered to UE is evaluated in a simple way. Moreover, a simple integration over the cell range allows to calculate the average throughput of the cell.

These results on the fluid model are valid for a constant inter site distance 2$R_c$. In the following section, we remind that we established \cite{ArxivFluid} a correspondence between a stochastic Poisson network and the Fluid network model.

\subsection{Comparison of the SINR CDF for the Poisson and Fluid models}
Since the CDF of the SINR characterizes the performance and the quality of service of wireless systems, we established a modified expression of the SINR given by the fluid model, which allows to calculate the CDF of SINR for the Poisson network reminded hereafter \cite{ArxivFluid}:
\begin{equation}
SINR^{modified}_{Fluid} = SINR_{Fluid}-(a \eta + b)
\label{fitting}
\end{equation}
which yields
\begin{equation}
CDF_{Poisson} \approx CDF^{modified}_{Fluid}
\label{cdffitting}
\end{equation}
where $a=$ 3 and $b=$ -6,
for a wide range of values of $\eta$ (ranging from 2.2 to 4.2, usual range for the path-loss exponent is comprised between 2.8 and 3.6)

We observed that for $\eta$ = 2.8, 3, 3.6, 3.8, the CDF of the SINR established by the \textit{modified Fluid model} and by the \textit{Poisson model} are very close: the differences between them are less than 0.4 dB \cite{ArxivFluid}.

\section{SINR calculation with Pathloss and Shadowing} \label{SINRCalculation}

We consider N interfering base station (BS), a mobile $u$ and its nearest $BS_0$.

\subsection{Impact of the pathloss on the SINR}

By considering expression  (\ref{SIRdist}), we can see that the calculation of the SINR of a user depends on its location.
We notice that in this case, users are connected to their best serving station which is also their nearest BS.

\subsection{Pathloss and Shadowing impact on the SINR}
\subsubsection{Propagation} 
Considering the power $P_j$ transmitted by the BS \textit{j}, the power $p_{j,u}$ received by a mobile \textit{u} 
can be written: 
\begin{equation} \label{propag}
p_{j,u}= P_{j}Kr_{j,u}^{-\eta} Y_{j,u},
\end{equation}
where $Y_{j,u}=10^{\frac{\xi_{j,u}}{10}}$ represents the shadowing effect. The term $Y_{j,u}$ is a lognormal random variable characterizing the random variations of the received power around a mean value.  $\xi_{j,u}$ is a \textit{Normal} distributed random variable (RV), with zero mean and standard deviation, $\sigma$, comprised between 0 and 10~dB. The term $P_{j}Kr_{j,u}^{-\eta}$, where K is a constant, represents the mean value of the received power at distance $r_{j,u}$ from the transmitter $(BS_j)$. The probability density function (PDF) of this slowly varying received power is given by 
\begin{equation} \label{aqteta}
p_Y(s)= \frac{1}{a\sigma s\sqrt{\pi}}\exp - \left(\frac{\ln(s)-am}{\sqrt{2}a\sigma} \right)^2
\end{equation}
where 
$a=\frac{\ln10}{10}$,
$m=\frac{1}{a}\ln(KP_jr^{-\eta}_{j,u})$ is the (logarithmic) received mean power expressed in decibels (dB), which is related to the path loss and
$ \sigma$ is the (logarithmic) standard deviation of the mean received signal due to the shadowing.

\subsubsection{SINR of a User connected to its nearest BS} \label{sinrshadowingnearest}
Considering the useful power $P_0$ transmitted by its nearest base station $BS_0$, the useful power $p_{0,u}$ received by a mobile \textit{u} connected to $BS_0$ 
can be written: 
\begin{equation} \label{propag2}
p_{0,u}= P_{0}Kr_{0,u}^{-\eta} Y_{0,u}.
\end{equation}
For the sake of simplicity, we now drop index $u$ and set $r_{0,u}=r$. The interferences received by $u$ coming from all the other base stations of the network are expressed by:
\begin{equation}
p_{ext} = \sum_{j=1}^N P_j K r_{j}^{- \eta} Y_j.
\end{equation}
The SINR at user $u$ is given by:
\begin{equation}
\gamma = \frac{P_{0}Kr^{-\eta} Y_{0}}{ \sum_{j=1}^N P_j K  r_{j}^{- \eta} Y_j+N_{th}}.
\end{equation}

\subsubsection{SINR of a User connected to its best serving BS} 
In this case, the user is connected to the BS which offers the highest useful signal. Therefore, the SINR at user $u$ is given by:
\begin{equation}\label{sinrshadowingbestserver}
\gamma = \frac{ \max_{j=0}^N (P_{j}Kr_j^{-\eta} Y_{j})}{ \sum_{j \neq j^*}^N P_j K  r_{j}^{- \eta} Y_j+N_{th}}.
\end{equation}
where $j^*$ is the BS which offers the highest useful signal to the user $u$.
We notice that in this case, due to shadowing, users are connected to a BS which is not necessary the nearest one.

\section{Shadowing Impact on the SINR}\label{shadowing}
In this section, we analyze the shadowing impact on the CDF on the SINR in a system modeled by a Poisson network. 

\subsection{Base stations distribution}
In order to place the base stations, we set the \textit{expected half inter site distance} as $R_c$. This hypothesis fixes the density of base stations $\rho_{BS}$. The values of $\rho_{BS}$ and of the studied surface $S_A$ give the Poissonian characteristic of the network: the number of BS is drawn according to a Poisson distribution of parameter $\rho_{BS} S_A$. The surface is chosen to obtain in average 50 stations in the area. It allows to have a significant number of cells, representative of a realistic zone covered by BS, and a significant number of interfering BS for the computation of the SINR. Those BS are then placed in the network, with no pairwise constraint (Fig. \ref{Poissonnetwork}): distances between neighboring base stations may be very low.

\subsection{Poisson network SINR computation}
The users are uniformly distributed on the whole area $S_A$. Then, the SINR of a UE is computed from its definition: the best serving BS gives the power of the received signal, and all the other stations generate interference. Several Monte Carlo simulations are run. At each run, the number and locations of the BS change, whereas the set of studied points (UE) is fixed. As a result, for the set of studied points, we obtain the corresponding SINR with different configurations of BS. Therefore, it becomes easy to compute the CDF of the SINR received by UE in this zone. Considering a toroidal shape of the network allows to consider it as virtually infinite with no ``edge effect'' for the computation of the SINR.  

\subsection*{Remark} 
We consider a model proposed in \cite{Vit95}. The Rayleigh fading, not considered in this model, may be indistinguishable from shadowing if the fading is sufficiently slow \cite{Vit95}, as for example if the mobile travels through a region of deep fades at a very slow speed.

\subsection{Cumulated Distributed Function of the SINR with Shadowing}\label{shadowingpoisson}

We consider a Poisson model network. And we calculate the CDF of the SINR, in the zone covered by this network, by taking into account the shadowing. The UE are connected to their best serving station, i.e. the BS which offers the best signal. The SINR is calculated by using (\ref{sinrshadowingbestserver}).

\begin{figure}[htbp]
\begin{center}
\includegraphics[width=1\linewidth]{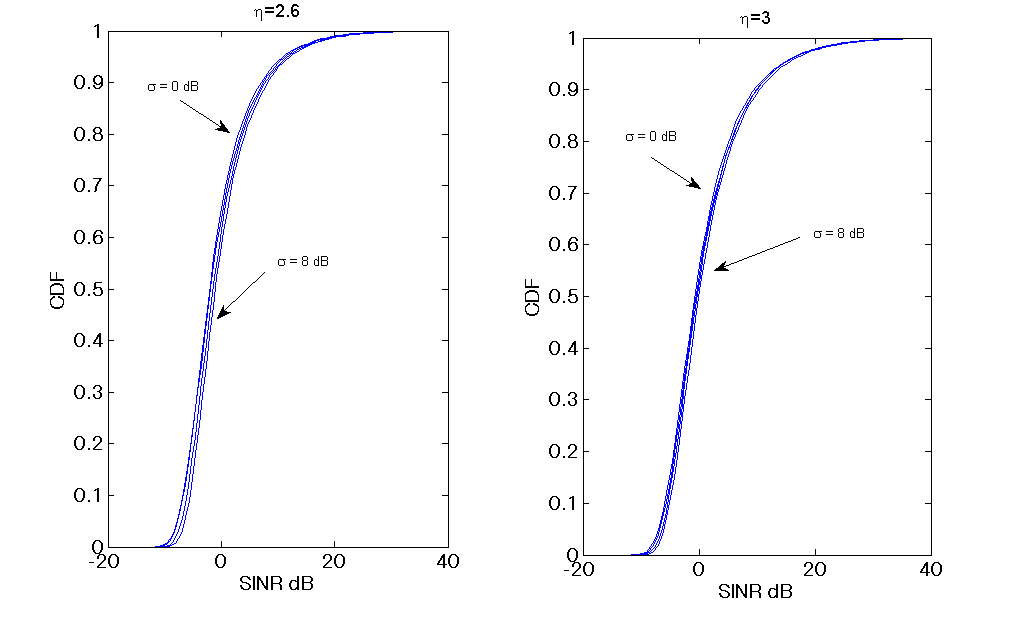}
\caption{CDF of the SINR with $\eta$ = 2.6 (left) and $\eta$ = 3 (right), for a Poisson model network and standard deviations of the shadowing values comprised between 0 dB (no shadowing) (curve in the left of each figure) and 8 dB (curve in the right of each figure). For readability, only the curves drawn with $\sigma$ = 0 and 8 dB are indicated in the figure. The curves between these two curves represent the cases $\sigma$ = 3 dB, 6 dB. They are in the line thickness}.
\label{poissonshadow26et3}
\end{center}
\end{figure}

\begin{figure}[htbp]
\begin{center}
\includegraphics[width=1\linewidth]{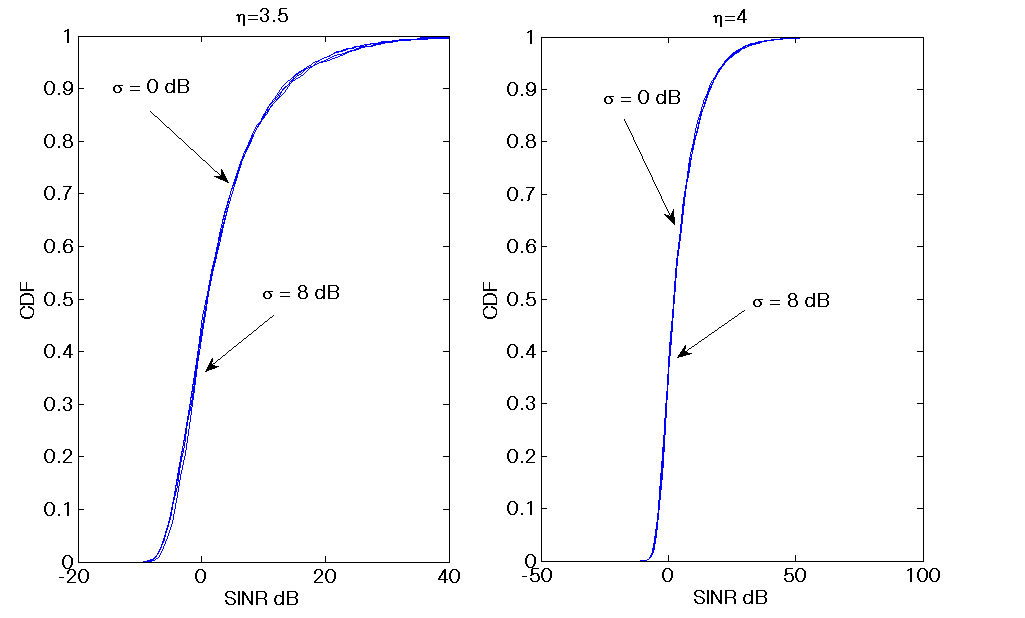}
\caption{CDF of the SINR with $\eta$ = 3.5 (left) and $\eta$ = 4 (right), for a Poisson model network and standard deviations of the shadowing values comprised between 0 dB (no shadowing) (curve in the left of each figure) and 8 dB (curve in the right of each figure). For readability, only the values of 0 and 8 dB are indicated in the figure. The curves between these two curves represent the cases 3 dB, 6 dB. They are in the line thickness}.
\label{poissonshadow35et4}
\end{center}
\end{figure}

The figures \ref{poissonshadow26et3} and \ref{poissonshadow35et4} show the CDF of the SINR for different values of the pathloss parameter $\eta$ and for standard deviation of values 0 dB (no shadowing), 3 dB, 6 dB, and 8 dB. It can be observed that for $\eta$  values of 3.5 and 4  (Fig. \ref{poissonshadow35et4}), the curves between 0 and 8 dB are indistinguishable. Only for $\eta$ = 2.6 (Fig. \ref{poissonshadow26et3}) it is hardly possible to distinguish between the different curves. 
These curves show that for typical values of the pathloss parameter $\eta$, comprised between 2.6 and 4, and values of the standard deviation of the shadowing ($\sigma \leq $ 6 dB) the impact of the shadowing is negligible : the difference $\delta$ between the CDFs with shadowing and without shadowing is less than 0.6 dB.
Let us notice that for $\eta= 2.6$, and for an outage of 5\%, the difference $\delta$, between the curve without shadowing and the curve with $\sigma$ = 8 dB, may reach 1.3 dB.

Therefore, the impact of the shadowing is low in terms of reachable throughput, coverage (probability of outage) and QoS.

\subsection{Explanation of these results}\label{shadowingexplanation1}

\begin{figure}[htbp]
\begin{center}
\includegraphics[width=1\linewidth]{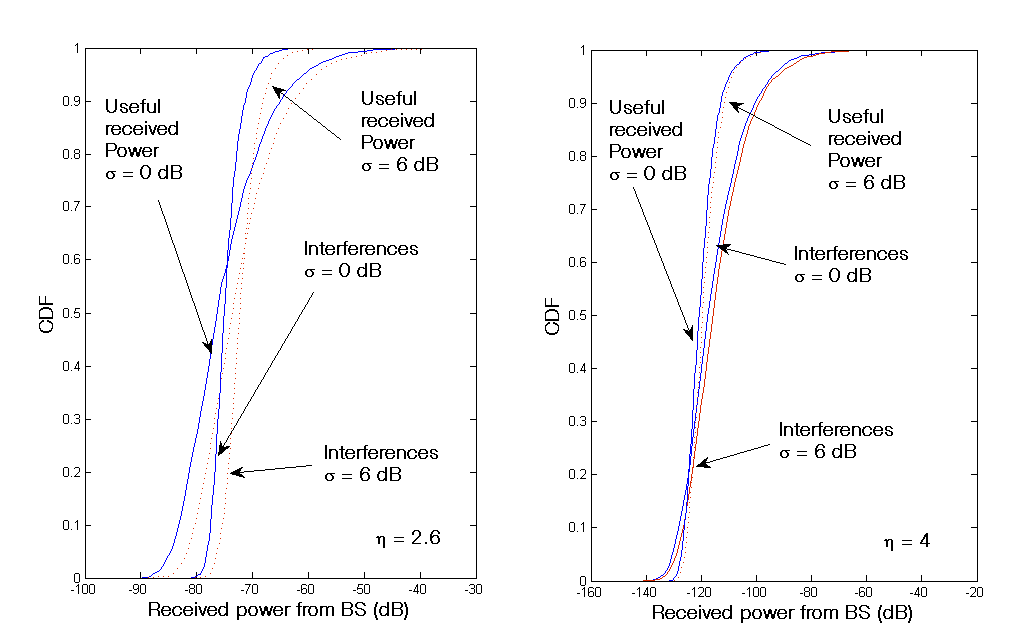}
\caption{CDF of the useful signal and the interferences, for $\eta$ = 2.6 (left) and $\eta$ = 4 (right), without shadowing ($\sigma$ = 0 dB, blue continuous curves) and with shadowing ($\sigma$ = 6 dB, red dotted curves)}.
\label{poissonBestInterf36et4}.
\end{center}
\end{figure}

In the aim to understand the phenomena which induce such curves, we analyzed the impact of the shadowing on the best useful received signal received by users, and on the interferences (Fig. \ref{poissonBestInterf36et4}).


These curves show that the CDF of the useful signal received by users is better with shadowing than without. It is due to the fact that if a user receives a low signal from its nearest BS, it has a non negligible probability to receive a better signal from another BS due to shadowing, and therefore to be connected to that BS than to be connected to its nearest BS (case without shadowing).
It can be observed that the interferences increase, too. And the increase of the interferences is at the same order as the increase of the useful signal. 
Therefore the two increases compensate each other.

\section{Impact on Wireless System Analysis}\label{shadowingimpactsimulation}
In the previous section, we established the CDF of the SINR in a system modeled by a Poisson network, and by taking into account the propagation in terms of pathloss and shadowing. We showed that the shadowing impact is very low for typical values of its standard deviation. Indeed, we showed that it is equivalent, in terms of performance and QoS, to consider the shadowing and a best serving BS policy, as to consider a nearest serving BS policy without taking into account the shadowing. Therefore, the analysis of performances, coverage/capacity and QoS of wireless networks can be done by taking into account the propagation in a very simple way: by only considering the pathloss.

Moreover, in Monte Carlo simulators, the implementation of the shadowing is particularly greedy in simulation time. For this reason also, it is interesting not to take it into account. 
Therefore a simple analytical wireless model which does not take into account the shadowing, is sufficient for most of the analysis. The CDF of the SINR can be calculated in a easy and quick way, by using the expressions (\ref{SIRfluid}), (\ref{fitting}) and (\ref{cdffitting}) given by the Fluid network model.


\section{Conclusion}\label{Conclusion}
In this paper, we analyzed the joint impact of the pathloss and the shadowing in Poisson wireless networks by considering a best serving policy for users.
We established that results are very close, in terms of performance and QoS, to the ones established by considering a nearest serving policy, without taking into account the shadowing. 
Therefore it becomes possible to analyze realistic wireless networks without considering the shadowing, in the standard range of values of propagation pathloss parameter $\eta$, and standard deviation of shadowing $\sigma$.
As a consequence the CDF of the SINR, established by Monte Carlo simulation for the Poisson model network is very close to the one calculated with the analytical expression of the SINR given by the Fluid model network when a linear function of the propagation parameter $\eta$ is applied. 
Therefore, the analysis of performance, outage probability, throughput, the radio resource allocation, the scheduling policy can be done in an easy and quick way, by using the expressions established with the Fluid Model network.

\section*{Acknowledgment}
This work was supported by the Seventh Framework Program for Research of the European Commission under grant number HARP-318489.


\end{document}